\documentclass[12pt]{article}
 \usepackage{putex, hyperref, epsfig}

 \newcommand{\beq}[1]{\begin{eqnarray}\label{#1}}
 \newcommand{\eeq}{\end{eqnarray}}

\begin{document}

 \institution{ITP-CAS}{Institute of Theoretical Physics, Beijing 100080, China}

 \title{Wake of color Fields in charged ${\cal N}=4$ SYM plasmas}

 \authors{Yi-hong Gao, Wei-shui Xu, and Ding-fang Zeng
 \footnote{ gaoyh@itp.ac.cn, wsxu@itp.ac.cn, and dfzeng@itp.ac.cn }}

 \abstract{The dissipative dynamics of a heavy quark passing through charged thermal
 plasmas of strongly coupled ${\cal N}=4$ super Yang-Mills theory is studied using AdS/CFT.
 We compute the linear response of the dilaton field to a test string in the rotating
 near-extremal D3 brane background, finding the momentum space profile of $\langle\textrm{tr}F^{2}\rangle$
 numerically. Our results naively support the wake picture discussed in hep-th/0605292, provided the
 rotation parameter is not too large.}

\date{June 2006}

\maketitle

\section{Introduction}

According to the AdS/CFT correspondence
\cite{Maldacena:1997re,Gubser:1998bc,Witten:1998qj,Witten:1998zw},
the physics of a heavy quark passing through finite temperature
 ${\cal N}=4$ super Yang-Mills plasmas may have the dual description
 in terms of a fundamental string in the background formed by a stack
 of near-extremal D3-branes. Such a subject has been studied in a number
 of recent papers
 \cite{Liu:2006ug,Herzog:2006gh,Buchel:2006bv,Gubser:2006bz,Herzog:2006se,Casalderrey-Solana:2006rq,Caceres:2006dj,FGM:2006bz,Elena:2006,LinFengli:2006,Avramis:2006,Armesto:2006},
 motivated by its possible connection with jet-quenching
 observed at RHIC \cite{Arsene:2004fa,Adcox:2004mh,Back:2004je,Adams:2005dq}
 in relativistic heavy ion collisions. This remarkable phenomenon can be
 understood as the strong energy loss of a high energy parton moving
 through the quark-gluon plasma; for some non-AdS/CFT based theoretical
 studies, see {\it e.g.}
 \cite{Greco:2003xt,Hwa:2004ng,Majumder:2004pt,Fries:2004hd,Armesto:2004pt,Casalderrey-Solana:2004qm,Ruppert:2005uz}.

In \cite{Herzog:2006gh}\cite{Gubser:2006bz}, the AdS/CFT duality was
applied to the computation of the drag force on a moving quark
received by a hot ${\cal N}=4$ SYM plasma. The background metric in
this dual description comes from near-extremal static D3-branes,
which is simply $AdS_{5}$-Schwarzschild$\times S^5$ in the
near-horizon limit and corresponds to a neutral quark-gluon plasma.
Such a computation was extended to $R$-charged ${\cal N}=4$ SYM
plasmas \cite{Herzog:2006se}\cite{Caceres:2006dj}, where the
background is taken to be the near-horizon geometry of rotating D3
branes, giving rise to the drag force exerted by the thermal plasma with a
non-vanishing chemical potential. However, both of these works
relied on the test string approximation; back-reaction of such a
string on the background was completely neglected. More recently,
the authors of \cite{FGM:2006bz} have computed linear responses of
the dilaton field to a test string in the
$AdS_{5}$-Schwarzschild$\times S^5$ background, which takes some
back-reaction effects into account. Their result is quite
interesting, exhibiting a recoil energy scale in some ranges and a
directional peak of gluons radiated from the heavy quark, in
consistent with the wake scenarios
\cite{Fries:2004hd,Ruppert:2005uz} of jet-quenching. With AdS/CFT,
this kind of computations should provide us useful information on
the energy loss of heavy quarks passing through the plasma.

In this note we wish to extend the result of \cite{FGM:2006bz} to
$R$-charged thermal plasmas, by studying the linear response of the
dilaton field to a test string in the rotating near-extremal D3
brane background. We shall treat the neutral plasma as a special
case of the charged ones, to see how the profile of
$\langle\textrm{tr}\,F^{2}\rangle$ changes and, in particular,
whether a wake form can be found in the presence of charges. For
simplicity, we will restrict ourselves to the case where only one
angular momentum parameter is non-vanishing
\cite{Russo:1998nc}\cite{CG:1999pr}. The near horizon geometry thus
reads \beq{}
ds^2&&\hspace{-5mm}=f^{-\frac{1}{2}}(-hdt^2+d\vec{x}^2)+f^{\frac{1}{2}}(\tilde{h}^{-1}dr^2
-\frac{2lr_0^2L^2}{r^4\Delta f}\sin^2\theta dtd\phi\nonumber\\
&&\hspace{30mm}
+r^2[\Delta d\theta^2+\tilde{\Delta}\sin^2\theta d\phi^2+\cos^2\theta d\Omega_3^2])
\label{spinD3metric}
\eeq
with
\beq{}
f&&\hspace{-5mm}=\frac{L^4}{r^4\Delta},\nonumber\\
\Delta&&\hspace{-5mm}=1+\frac{l^2}{r^2}\cos^2\theta,\nonumber\\
\tilde{\Delta}&&\hspace{-5mm}=1+\frac{l^2}{r^2}+\frac{r_0^4l^2\sin^2\theta}{r^6\Delta f},\nonumber\\
h&&\hspace{-5mm}=\frac{1}{\Delta}(1+\frac{l^2}{r^2}\cos^2\theta-\frac{r_0^4}{r^4}),\nonumber\\
\tilde{h}&&\hspace{-5mm}=\frac{1}{\Delta}(1+\frac{l^2}{r^2}-\frac{r_0^4}{r^4})
\eeq and $L^{4}= g_{YM}^{2}N\alpha^{\prime 2}$. For later
convenience, let us introduce two symbols $h_1$, $h_2$
\begin{subequations}
\beq{}
h_1=\left[\frac{1}{2}(\sqrt{l^4+4r_0^4}-l^2)\right]^{\frac{1}{2}}
\label{h1define}
\\
h_2=\left[\frac{1}{2}(\sqrt{l^4+4r_0^4}+l^2)\right]^{\frac{1}{2}}
\label{h2define}
\eeq
\end{subequations}
where $h_1$ is the event horizon. We will consider a classical
string placed in the background (\ref{spinD3metric}), with one end
attached to the boundary of the AdS space moving in a constant
velocity $v$, and the other end attached to the horizon of the black
hole at the center of the AdS space. Our purpose here is to compute
the linear response of the dilaton in the background metric
(\ref{spinD3metric}), paying particular attention to the energy
scale at which dissipation occurs as well as the existence of
directional peaks.

This paper is organized as follows: In section 2 we recall some
relevant formulae of \cite{Herzog:2006se,Caceres:2006dj}. In section
3, we derive a set of differential equations governing linear
perturbations of the dilaton field, following \cite{FGM:2006bz}
closely. Section 4 then applies the relaxation method
\cite{WHpress:93bz} to the boundary value problem, and presents our
numerical results with some discussions. As we will see, in the case
of $l=0$, our results agree with those given in \cite{FGM:2006bz},
where a different numerical method was used.

\section{Test String Solutions}

Let us begin with the string configuration
\beq{}
X^0=\tau,\ X^r=\sigma,\ X^1=v\tau+\xi(r),\ \theta=\theta(r)
\label{stringSolution}
\eeq
The function $\xi(r)$ in ({\ref{stringSolution}) is determined by
\beq{}
\xi^\prime=\frac{vr_0^2}{L^2}\frac{f}{h\tilde{h}}\sqrt{1+\Delta\tilde{h}r^2\theta^{\prime2}}
\label{dxidr}
\eeq
and $\theta(r)$ obeys a complicated equation \cite{Herzog:2006se,Caceres:2006dj},
which in generical (for $l\neq 0$) does not allow constant solutions. However,
there are two special cases where $\theta(r)$ can have $r$-independent solutions,
provided $\theta\rightarrow 0\ or\ \frac{\pi}{2}$ as $r\rightarrow\infty$. In the
first case the string is parallel to the rotation axis, while in the second case
the string lies in the plane perpendicular to the rotation axis. In the present
paper we will consider the first case only. This string configuration (with $\theta\equiv0$),
though special, is nevertheless non-trivial, since it could be used to study the
physical effects of the SYM plasma charge on the test string.

Inserting $\theta\equiv 0$ into (\ref{dxidr}) gives
\beq{}
\xi^\prime=\left.\frac{vr_0^2}{L^2}\frac{f}{h}\right|_{\theta=0}=\frac{vr_0^2L^2}{r^4+r^2l^2-r_0^4}.
\label{case1Dxi}
\eeq
One may write down an explicit expression for $\xi(r)$ by integrating out the
above equation. With the definition (\ref{h1define})-(\ref{h2define}), it is
easy to derive:
\beq{}
\xi(r)=\frac{vr_0^2L^2}{\sqrt{l^4+4r_0^4}}\left[
\frac{\pi}{2h_2}-\frac{1}{h_1}\ln\frac{\sqrt{r+h_1}}{\sqrt{r-h_1}}-\frac{1}{h_2}\arctan\frac{r}{h_2}\right].
\label{case1xi}
\eeq

Now, to compare this with \cite{Gubser:2006bz}\cite{FGM:2006bz}, we set
\beq{}
z_H=\frac{r_0^2L^2}{h_1\sqrt{l^4/4+r_0^4}}
\eeq
so that
\beq{}
\xi(r)=\frac{vz_H}{4}\left[
\frac{\pi h_1}{h_2}-\ln\frac{r+h_1}{r-h_1}-\frac{2h_1}{h_2}\arctan\frac{r}{h_2}\right]
\label{case1xiGubConvention}.
\eeq
Note that Eq.(\ref{case1xiGubConvention}) under the limit $l\rightarrow 0$ will
become identical to the string configuration
$$
\xi(r)=\frac{L^{2}v}{2r_{0}}\left(\frac{\pi}{2}-\arctan\frac{r}{r_0}-\log\sqrt{\frac{r+r_0}{r-r_0}}\right)
$$
studied in \cite{Gubser:2006bz}\cite{FGM:2006bz}. This could be expected, since in that
limiting case the metric (\ref{spinD3metric}) reduces exactly to $AdS_5$-Schwarzschild.
Now for generic $l\neq 0$, the effective 5-dimensional background felt by the string
parallel to the rotation axis also gets simplified. Actually, taking $\theta\equiv 0$
and freezing all the angular degrees of freedom in (\ref{spinD3metric}) yields:
\beq{}
&&\hspace{-5mm}ds^2=-f^{-1/2}hdt^2+f^{1/2}h^{-1}dr^2+f^{-1/2}d\vec{x}^2,\nonumber\\
&&\hspace{-5mm}f=\frac{L^4}{r^4+r^2l^2},\ h=1-\frac{r_0^4}{r^4+r^2l^2}.
\label{metricD5}
\eeq

\section{Linear Responses of the Dilaton}

 We turn now to back-reaction of the test string to the background, following
 the same routine of \cite{FGM:2006bz}. Consider the response of the dilaton field
 $\phi$ to the test string, where $\phi$ has non-trivial solutions determined by
 minimizing the action
 \beq{}
 S&&\hspace{-5mm}=-\frac{1}{4\kappa_5^2}\int dx^5\sqrt{-G}(\partial\phi)^2
 -\frac{1}{2\pi\alpha^\prime}\int_Md^2\sigma
 e^{\phi/2}\sqrt{-\gamma}
 \eeq
 with
 \beq{}
 \gamma_{ab}=\partial_aX^\mu\partial_bX^\nu G_{\mu\nu},
 \ \kappa_5^2=8\pi G_5=\frac{4\pi^2L^3}{N^2}.
 \eeq

 To establish the equation of motion for $\phi$, it is convenient to express the
 action as a single volume integral \cite{FGM:2006bz}
 \beq{}
 S&&\hspace{-5mm}=-\frac{1}{4\kappa_5^2}
 \int dx^5\sqrt{-G}\left[(\partial\phi)^2
 +\frac{4\kappa_5^2}{2\pi\alpha^\prime}\int_Md^2\sigma
 e^{\phi/2}\frac{\sqrt{-\gamma}}{\sqrt{-G}}\delta^5(x^\mu-X^\mu(\sigma))\right].
 \label{basicAction}
 \eeq
 This gives rise to the linearized equation of motion describing the
 response of the dilaton to the test string:
 \beq{}
 \Box\phi=\frac{1}{\sqrt{-G}}\partial_\mu\sqrt{-G}G^{\mu\nu}\partial_\nu\phi
 =\frac{\kappa_5^2}{2\pi\alpha^\prime}\int_Md^2\sigma
 e^{\phi/2}\frac{\sqrt{-\gamma}}{\sqrt{-G}}\delta^5(x^\mu-X^\mu(\sigma))
 \equiv J\label{Jdefinition1}
 \eeq
Substituting (\ref{stringSolution}), (\ref{case1Dxi}), (\ref{case1xi}) into the
above definition of $J$ and calculating $\sqrt{-\gamma}$ as well as $\sqrt{-G}$
in terms of the metric components, one finds
\beq{}
J&&\hspace{-5mm}=\frac{\kappa_5^2}{2\pi\alpha^\prime}
\frac{\sqrt{-\gamma}}{\sqrt{-G}}\delta(x^1-X^1(t,r))\delta(x^2-X^2)\delta(x^3-X^3)
\label{Jdefinition2}
\\
\sqrt{-G}&&\hspace{-5mm}=\sqrt{-G_{tt}G_{rr}G^3_{xx}}=G_{xx}^{3/2}(\theta\equiv 0)\nonumber\\
\sqrt{-\gamma}&&\hspace{-5mm}=\sqrt{(h-v^2)\tilde{h}^{-1}+hf^{-1}\xi^{\prime2}+\Delta(h-v^2)r^2\theta^{\prime2}}=\sqrt{1-v^2}
\eeq

Suppose, just as in \cite{FGM:2006bz}, that $\phi$ depends on $x^1$
and $t$ only through the combination $x^1-vt$. Also, notice that in
our string configuration, $\theta\equiv 0$. Hence we can simplify
Eq.(\ref{Jdefinition1}) to
\beq{} &&\hspace{-5mm}\partial_r
G_{xx}^{3/2}G^{rr}\partial_r\phi
+G_{xx}^{3/2}\left[(G^{xx}+v^2G^{tt})\partial_1^2
+G^{xx}(\partial_2^2+\partial_3^2)\right]\phi\nonumber\\
&&\hspace{25mm}=\frac{\kappa_5^2\sqrt{1-v^2}}{2\pi\alpha^\prime}
\delta(x^1-vt-\xi(r))\delta(x^2-X^2)\delta(x^3-X^3) \eeq Now, going
to the momentum space \beq{}
\phi(t,r,\vec{x})=\int\frac{d^{3}\vec{k}}{(2\pi)^3}e^{i[k_1(x^1-vt)+ik_2x^2+ik_3x^3]}\phi_k(r),
\quad \tilde{\phi}_k(r)\equiv\phi_k(r)\frac{2\pi\alpha^\prime
}{\kappa_5^2\sqrt{(1-v^2)}} \eeq the above equation is transformed
into \beq{}
&&\hspace{-5mm}\partial_r[f^{-\frac{5}{4}}h(r)\partial_r\tilde{\phi}_k(r)]
-f(r)^{-\frac{1}{4}}\left[[1-\frac{v^2}{h(r)}]k_1^2
+k_\bot^2\right]\tilde{\phi}_k(r)= e^{-ik_1\xi(r)},
\label{simplifiedBasicEq} \eeq here $f$ and $g$ are given in
(\ref{metricD5}).

It seems not possible to solve Eq.(\ref{simplifiedBasicEq}) analytically for the full
range of $r$, but nevertheless we can find explicit solutions in two asymptotical regions.
The first one is in the near-horizon region, where $r\sim h_{1}$ and therefore
\beq{}
f(r)&&\hspace{-5mm}\rightarrow\frac{L^4}{r_0^4}\nonumber\\
h(r)&&\hspace{-5mm}\rightarrow\frac{2h_1(h_1^2+h_2^2)(r-h_1)}{r_0^4}\equiv h_{0}(r)\nonumber\\
\xi(r)&&\hspace{-5mm}\rightarrow\frac{vz_H}{4}\left[ \frac{\pi
h_1}{h_2}-\ln\frac{2h_1}{r-h_1}-\frac{2h_1}{h_2}\arctan\frac{h_1}{h_2}\right]
\label{case1xiNHlimit} \eeq In this limit (\ref{simplifiedBasicEq})
reduces to \beq{} &&\hspace{-5mm}\frac{r_0^5h_{0}}{L^5}
\partial_r(r-h_1)\partial_r\tilde{\phi}_k(r)
+\frac{v^2r_0}{Lh_{0}(r-h_1)}k_1^2\tilde{\phi}_k(r)\nonumber\\
&&\hspace{15mm}=\exp\left[\frac{-ivk_1z_H}{4}\left(
\frac{\pi h_1}{h_2}-\ln\frac{2h_1}{r-h_1}-\frac{2h_1}{h_2}\arctan\frac{h_1}{h_2}\right)\right].
\eeq
Now let
\beq{}
Y=\ln\frac{r-h_1}{h_1},\ P&&\hspace{-5mm}=\frac{\pi h_1}{h_2}-\ln2-\frac{2h_1}{h_2}\arctan\frac{h_1}{h_2},
\eeq
we can rewrite the above differential equation as
 \beq{}
\partial_Y^2\tilde{\phi}_k(Y)+\left[\frac{vk_1z_H}{4}\right]^2\tilde{\phi}_k=\frac{z_Hh_1L^3}{4r_0^3}e^Ye^{-ivk_1z_H(Y+P)/4}
\label{simplifiedNHeq} \eeq The general solution of
Eq.(\ref{simplifiedNHeq}) thus takes the form \beq{}
\tilde{\phi}_{k,NH}(r)=\frac{z_Hh_1L^3}{4r_0^3}\frac{e^Ye^{-ivk_1z_H(Y+P)/4}}{1-ivk_1z_H/2}+C_k^{+}e^{ivk_1z_HY/4}+C_k^{-}e^{-ivk_1z_HY/4}
\label{NearHorizonSol} \eeq where $C_k^{\pm}$ are arbitrary
constants. In this near horizon region, we require $\phi$ depend on
$t$ and $Y$ through the combination $t+vk_1z_HY/4$ only, so we can
set $C_k^{+}=0$. Physically this means that we accept the infalling
solution while reject the outgoing one.

Next we consider asymptotical solutions in region near the (AdS) boundary, where
\beq{}
r\rightarrow \infty,
\ f^{-\frac{1}{4}}\rightarrow \frac{r}{L},
\ h\rightarrow 1
\eeq
In that region, Eq.(\ref{simplifiedBasicEq}) becomes
\beq{}
\frac{1}{r}\partial_rr^5\partial_r\tilde{\phi}_k(r)=\frac{L^5}{r}
\eeq
whose general solution has the form:
\beq{}
\tilde{\phi}_{k,NB}(r)=-\frac{1}{3}L^5r^{-3}+A_k+B_kr^{-4}\label{NearBoundarySol}
\eeq
The constant $A_k$ should be set to zero, as there are no deformations in the dilaton
Lagrangian. Physically we will be interested in $B_k$, since according to AdS/CFT,
this quantity is directly related to the vacuum expectation value of the operator
${\cal O}_{F^2}\sim {\rm tr}F^{2}$ coupled to the dilaton. Actually, using the AdS/CFT
dictionary we can write the VEV as
\beq{}
\langle{\cal O}_{F^2}(t,\vec{x})\rangle=\frac{1}{2\kappa_5^2}
\textrm{lim}_{r\rightarrow\infty}\sqrt{-g}g^{rr}\partial_r\phi
=-\frac{1}{2\kappa_5^2}\textrm{lim}_{r\rightarrow\infty}\frac{r^5}{L^5}\partial_r\phi(t,\vec{x},r)
\label{correspondence}
\eeq
where the metric (\ref{metricD5}) has been applied. Note that the above limit does not exist
due to the first term in Eq.(\ref{NearBoundarySol}). Since such a term does not depend on $\vec{k}$,
transforming it into real space will lead to a delta function supported at the quark location,
which should be subtracted \cite{FGM:2006bz}. The correspondence between $\langle{\cal O}_{F^2}\rangle$
and $\phi(t,\vec{x},r)$ in (\ref{correspondence}) should now be understood as a relation after the
delta-function subtraction. We thus find:
\beq{}
\langle\mathcal{O}_{F^2}(t,\vec{x})\rangle=-\frac{\sqrt{1-v^2}}{4\pi\alpha^\prime L^5}\int
\frac{d^{3}\vec{k}}{(2\pi)^3}e^{ik_1(x^1-vt)+ik_2x^2+ik_3x^3}B_k
\label{correspondenceConcrete}.
\eeq

Now, as in the neutral plasma case \cite{FGM:2006bz}, further subtraction is needed in order to
separate the dissipative dynamics from the near field contributions of the quark. One expects
that these near field contributions correspond to the string hanging straight down in $AdS_{5}$.
Thus, when $v=0$, we can apply the method of \cite{Danielsson:1998wt} directly to derive
\beq{}
\langle\mathcal{O}_{F^2}(t,x)\rangle^{\rm near\;field}\;=\;\frac{1}{16\pi^2}\frac{\sqrt{Ng_{YM}^2}}{|\vec{x}|^4}.
\label{nearFieldContributionBFT}
\eeq
Comparing this with (\ref{correspondenceConcrete}) for $v=0$, one gets
\beq{}
B_{k}^{\rm near\;field}=\frac{\pi \alpha^\prime L^5\sqrt{Ng_{YM}^2}}{16}\sqrt{k_1^2+k_{\bot}^2}.
\label{nearFieldContributionSF}
\eeq
In the case of $v\neq 0$, the near field contributions are obtainable through a Lorentz boost to
the string configuration. The result reads
\beq{}
B_{k}^{\rm near\;field}=\frac{\pi L^7}{16}\sqrt{(1-v^2)k_1^2+k_{\bot}^2},
\label{nearFieldContribution}
\eeq
which will be subtracted from the numerical values of $B_{k}$ computed in the next section.

\section{Numerical Results and Discussions}

We are now in a position to solve the following boundary value problem numerically:
\begin{subequations}
\beq{}
&&\hspace{-5mm}\partial_r[f^{-\frac{5}{4}}h(r)\partial_r\tilde{\phi}_k(r)]
-f(r)^{-\frac{1}{4}}\left[[1-\frac{v^2}{h(r)}]k_1^2
+k_\bot^2\right]\tilde{\phi}_k(r)=
e^{-ik_1\xi(r)}
\label{mspaceBasicEq2}\\
&&\hspace{-5mm}
\tilde{\phi}_{k}(r)\stackrel{r\sim h_{1}}{\longrightarrow}\frac{z_Hh_1L^3}{4r_0^3}\frac{e^Ye^{-ivk_1z_H(Y+P)/4}}{1-ivk_1z_H/2}+C_k^{-}e^{-ivk_1z_HY/4}
\equiv\tilde{\phi}_{k,NH1}+\tilde{\phi}_{k,NH2}
\label{NearHorizonSo2}\\
&&\hspace{-5mm}
\tilde{\phi}_{k}(r)\stackrel{r\sim \infty}{\longrightarrow}-\frac{1}{3}L^5r^{-3}+B_kr^{-4}
\equiv\tilde{\phi}_{k,NB1}+\tilde{\phi}_{k,NB2}
\label{NearBoundarySo2}
\eeq\label{boundaryProblem}
\end{subequations}
To this end, we need to implement the boundary conditions at $r\sim h_{1}$ and $r\sim\infty$, and
this can be done by introducing two Wronskians, $W_{NH}(r)$ and $W_{NB}(r)$, to measure the differences
between our numerically evaluated $\tilde{\phi}_{k}(r)$ and the asymptotical solutions (\ref{NearHorizonSol}),
(\ref{NearBoundarySol}) found in the last section. One thus defines, following \cite{FGM:2006bz},
\beq{}
W_{NH}(r)=(\tilde{\phi}_k-\tilde{\phi}_{k,NH1})\tilde{\phi}_{k,NH2}^\prime
-(\tilde{\phi}_k^\prime-\tilde{\phi}_{k,NH1}^\prime)\tilde{\phi}_{k,NH2}
\nonumber\\
W_{NB}(r)=(\tilde{\phi}_k-\tilde{\phi}_{k,NB1})\tilde{\phi}_{k,NB2}^\prime
-(\tilde{\phi}_k^\prime-\tilde{\phi}_{k,NB1}^\prime)\tilde{\phi}_{k,NB2}
\eeq
The boundary conditions can then be imposed properly by the requirements $W_{NH}(r)=0$ at $r\sim h_{1}$ and $W_{NB}(r)=0$ at $r\sim\infty$.

The numerical recipes we used in solving the boundary value problem
(\ref{boundaryProblem}) is the standard relaxation method
\cite{WHpress:93bz}. In programming, we changed the variable
$r\rightarrow y\equiv h_1/r$, so that the integration range becomes
$y\in(0,1)$, where $y\sim 1$ corresponds to the horizon and $y\sim
0$ to the $AdS_{5}$ boundary. Since the solution has oscillating
behavior at $y\approx 1$, we divided the the integration region into
four intervals, $(1,0.9]$, $(0.9,0.7]$, $(0.7,0.4]$ and $(0.4,0)$,
and then divided the first, second, third, and forth intervals into
4096, 1024, 256, and 64 integration steps, respectively, in order to
reach as high precision as possible while keeping CPU time in an
acceptable range. We also checked that if the whole integration
region is divided into 50000 steps uniformly, one can get almost the
same results and, of course, this costs more CPU time.

Since Eq.(\ref{mspaceBasicEq2}) becomes singular as $r\rightarrow h_{1}$ or $r\rightarrow\infty$,
we have to set $W_{NH}=0$ at a point very close to the horizon, and set $W_{NB}=0$ at a large but
finite value of $r$. The conditions imposed in our computations are
\beq{}
\left.W_{NH}\right|_{r=1.001h_1}=0,\quad \left.W_{NB}\right|_{r=700h_1}=0.
\label{bc}
\eeq

\begin{figure}[h]
\includegraphics[clip, scale=0.47, bb=45 177 543 788]{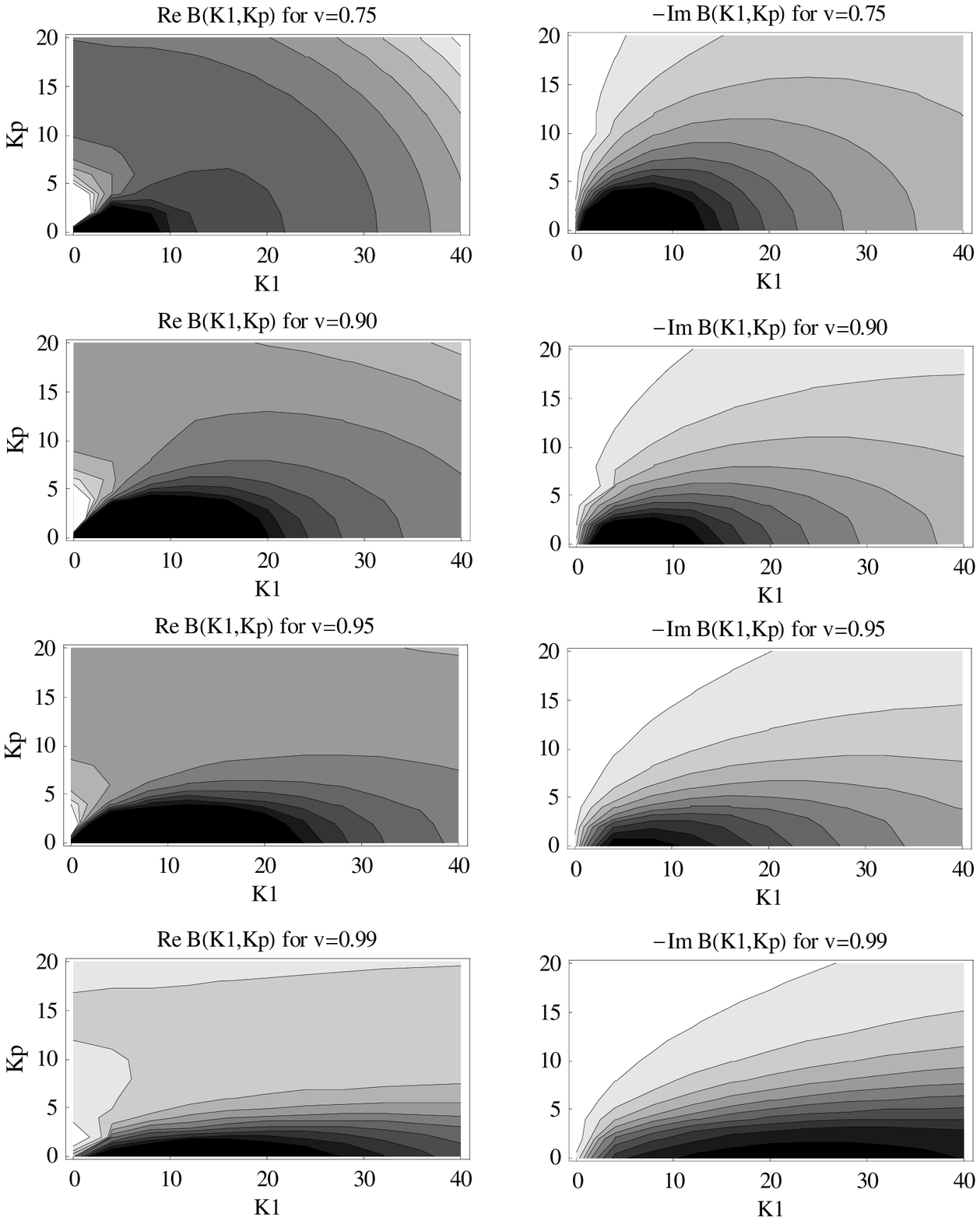}
\includegraphics[clip, scale=0.47, bb=45 177 543 788]{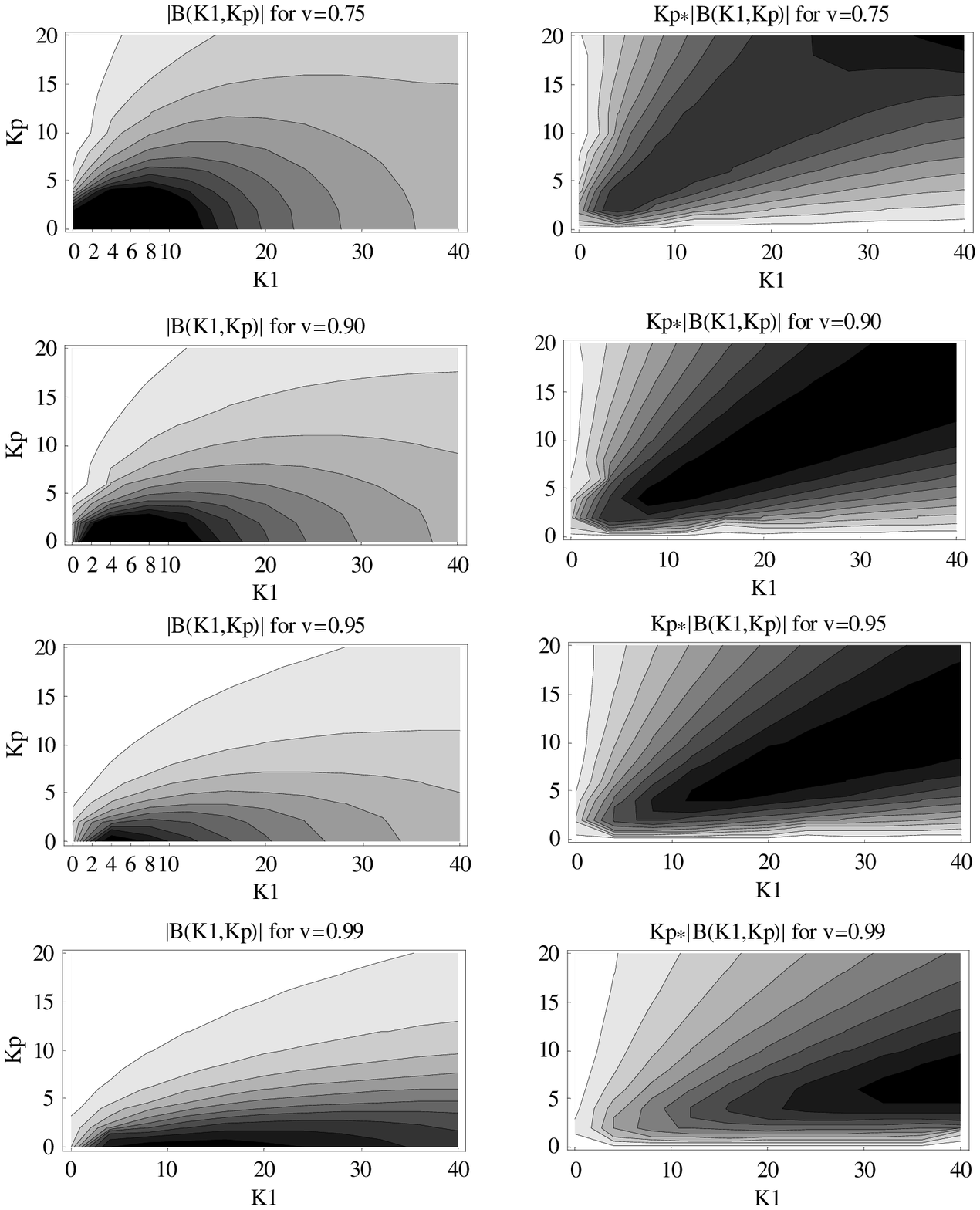}
\caption{Distributions of equi-value lines of
$\textrm{Re}B(K_1,K_\bot)$, $-\textrm{Im}B(K_1,K_\bot)$,
$|B(K_1,K_\bot)|$, and $K_\bot|B(K_1,K_\bot)|$ in the momentum plane
$(K_{1},K_{\bot})$, for angular momentum $l=0$, at speeds
$v=0.75,0.90,0.95,0.99$. We have subtracted the near field
contribution (\ref{nearFieldContribution}) from
$B(K_{1},K_{\bot})$.} \label{spin00Fig}
\end{figure}
\begin{figure}[h]
\includegraphics[clip, scale=0.47, bb=47 198 525 788]{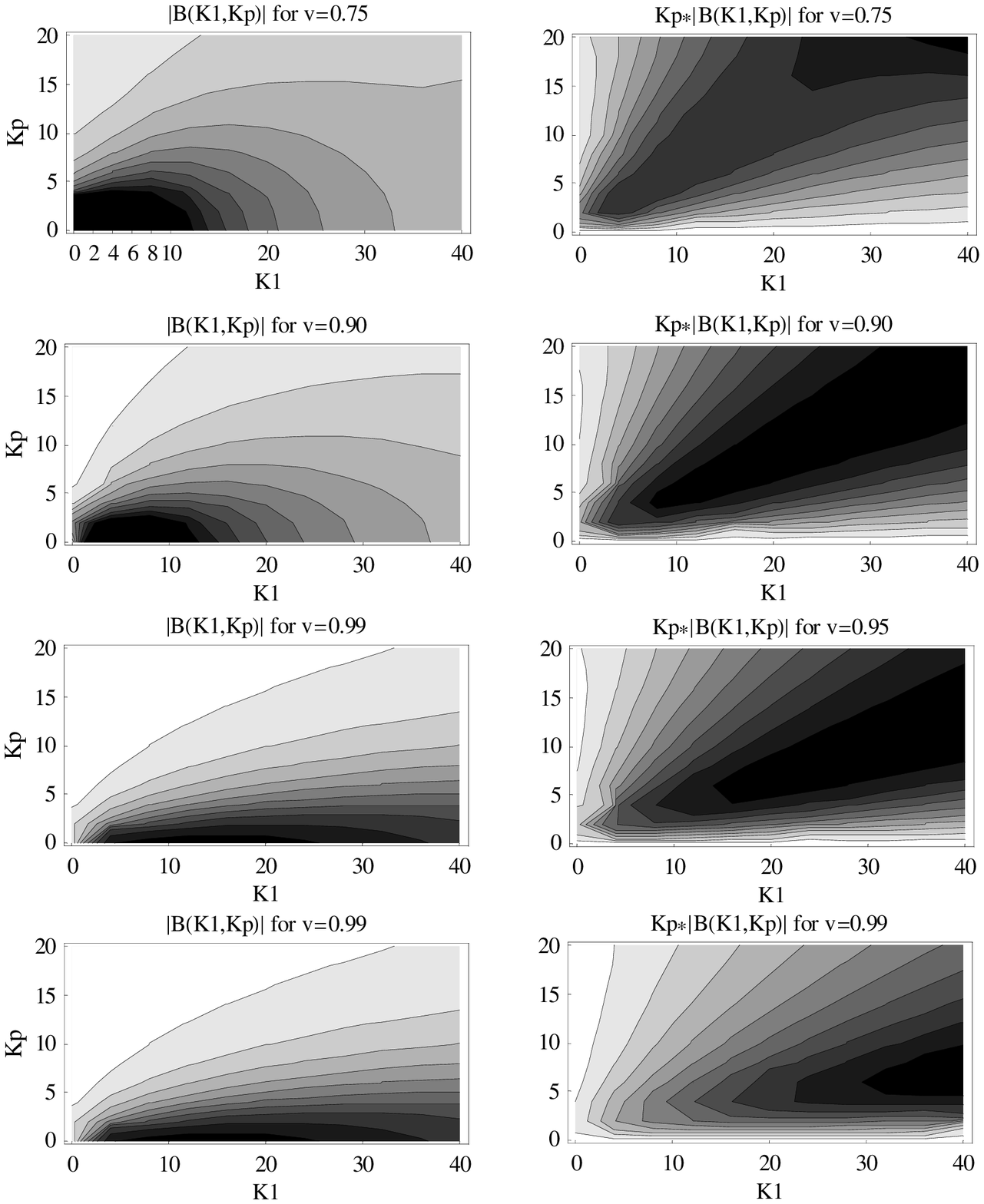}
\includegraphics[clip, scale=0.47, bb=47 198 525 788]{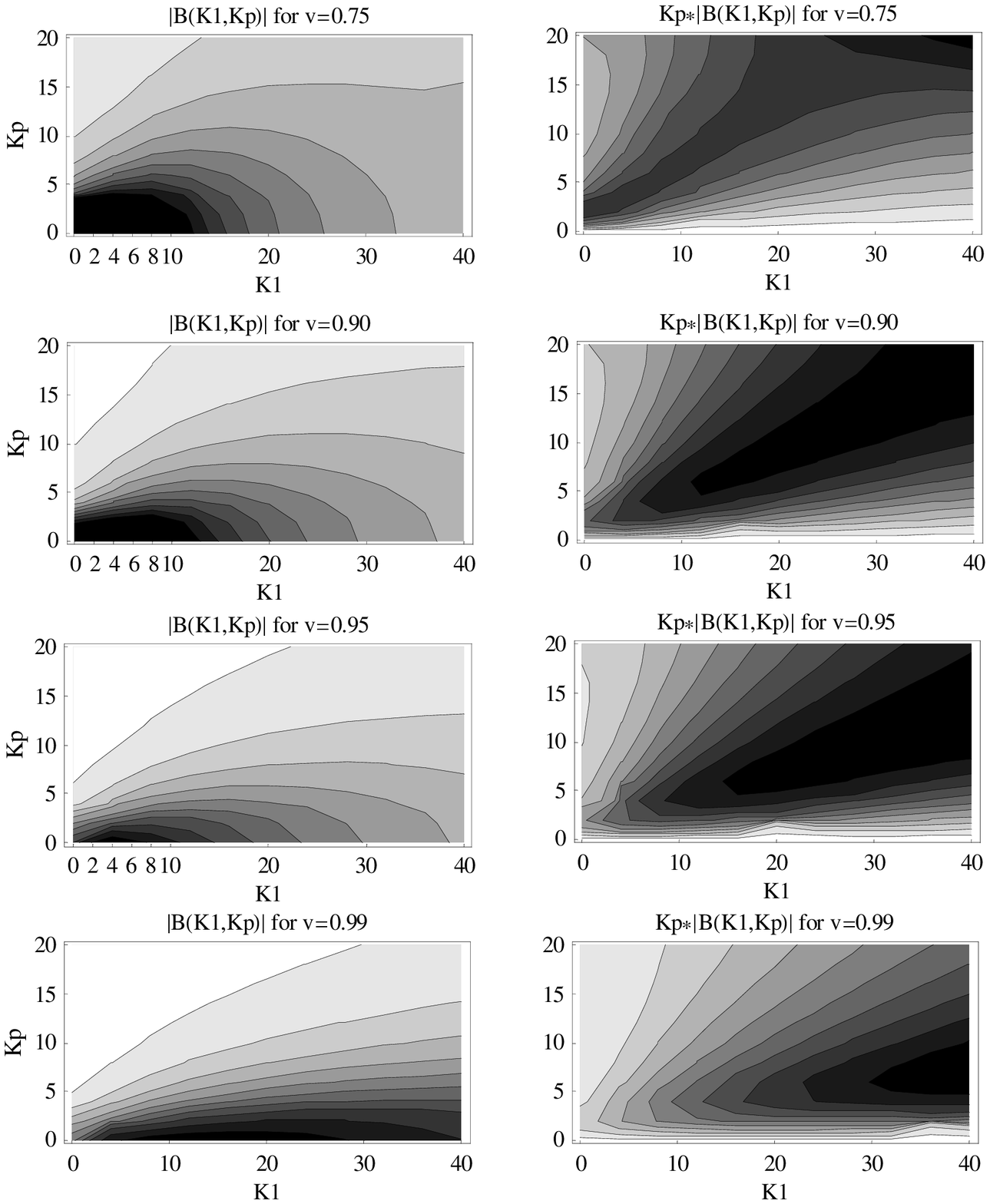}
\caption{Distributions of equi-value lines of $|B(K_1,K_\bot)|$ and
$K_\bot|B(K_1,K_\bot)|$ in the plane $(K_{1},K_{\bot})$, for angular
momentum $l=0.5$ and $1$, at speeds $v=0.75,0.90,0.95,0.99$. We have
subtracted the near field contribution (\ref{nearFieldContribution})
from $B(K_{1},K_{\bot})$. The left eight ones are plots of
$|B(K_1,K_\bot)|$ and $K_\bot|B(K_1,K_\bot)|$ for $l=0.5h_1$, and the
right eight ones are the corresponding plots for $l=1.0h_1$.}
\label{spin0510Fig}
\end{figure}

We depict our numerical results in
Fig.\ref{spin00Fig}--\ref{spin0510Fig}, where equi-value lines of
$\textrm{Re}B_{k}$, $-\textrm{Im} B_{k}$, $|B_{k}|$ {\it etc.}, for
some different values of $v$ and $l$, are plotted in the momentum
plane $k=(K_{1},K_{\bot})$. In all the plots the near field
contributions (\ref{nearFieldContribution}) have been subtracted. We
adopt the convention of \cite{FGM:2006bz}, representing values
closest to zero by white regions, and representing the most positive values by
black regions. To compare our results with those of
\cite{FGM:2006bz}, we fixed the energy scale explicitly, by setting
the temperature of the plasma \beq{} T=\frac{h_1}{2\pi
L^2r_0^2}\sqrt{l^4+4r_0^2} \eeq to be
$T=1/\pi\,\textrm{GeV}=318\textrm{MeV}$, in accordance with the
choice made in \cite{FGM:2006bz}. Thus, the wave numbers $k$
displayed here are measured by $\textrm{Gev}/c$. Apart from this,
the rotation parameter $l$ is shown in units of $h_1$, namely, when
we say $l=0.5$, we actually mean $l=0.5h_1$.

Fig.\ref{spin00Fig} shows our results of $B_{k}=B(K_{1},K_{\bot})$
in the special case $l=0$. This corresponds to the neutral plasma
studied in \cite{FGM:2006bz}. The first two columns contain plots of
$\textrm{Re}B(K_1,K_\bot)$ and $-\textrm{Im}B(K_1,K_\bot)$ at
$v=0.75,0.90,0.95,0.99$, while the third and forth show the
corresponding plots of $|B(K_1,K_\bot)|$ and
$K_\bot|B(K_1,K_\bot)|$, respectively. Comparing these with the
plots given in \cite{FGM:2006bz}, one sees that the basic features
are the same, both exhibiting a directionally peaked structure in
$K_\bot|B(K_1,K_\bot)|$ and a possible range of the recoil energy.
For example, focusing on the third line, third column of
Fig.\ref{spin00Fig}, we find that $|B(K_1,K_\bot)|$ (for $v=0.95$)
is peaked at
$$K_\bot\approx0,\quad 3\leq K_1\leq 7.2\;\;\textrm{Gev}/c,
$$
which indicates that the recoil energy $E_{r}$ is in the range
$1.5\leq E_{r}\leq 3.6$ Gev, less than the value
$$
E_{f}=\frac{1+v^{2}}{1-v^{2}}\,T=6.2\,\textrm{Gev}\quad\quad\quad\hbox{(for $v=0.95$)}
$$
predicted in the corresponding free field theory by a factor of a few. This agrees perfectly with the numerical result of \cite{FGM:2006bz}.

Now we turn to the case $l\neq 0$. When $l=0.5$ and $1$, our results
for $B(K_1,K_\bot)$ (at speeds $v=$ 0.75, 0.90, 0.95, 0.99) are
displayed in Fig.\ref{spin0510Fig}. The first and third columns give
the plots of $|B(K_1,K_\bot)|$ for $l=0.5$ and $l=1$, respectively. From
these plots, one sees that there may exist a finite range of $K_\bot$ in
which $|B(K_1,K_\bot)|$ is peaked, just as in the $l=0$ case. Taking the
plot for $l=1$ at $v=0.95$ as an example (the one placed at the third line, third column of Fig.\ref{spin0510Fig})), we find that $|B(K_1,K_\bot)|$ has a peak within
$$K_\bot\approx0,\quad 2\leq K_1\leq 7.6\;\;\textrm{Gev}/c,
$$
so that the recoil energy $E_{r}$ is roughly in the range $1\leq
E_{r}\leq 3.8$ Gev, which is slightly larger than the result found in the
$l=0$ case. That the range of $E_{r}$ (in particular its upper bound
value) becomes larger when $l$ increases seems to be a generic
phenomenon in our numerical computations. A possible implication is
that the more is the charge carried by the plasma, the more energy
of the quark would be dissipated away by radiation of gluons. This
should not be taken too seriously, however, since our numerical
results may contain more errors as $l$ becomes larger; see the
discussion for $l=2$ below.

The forth column of Fig.\ref{spin0510Fig} shows plots of
$K_\bot|B(K_1,K_\bot)|$ for $l=1$, at speeds $v=$ 0.75, 0.90, 0.95,
and 0.99. Again, we can clearly see a directionally peaked structure
in these plots, much resembling the $l=0$ case. We have also
displayed plots of $K_\bot|B(K_1,K_\bot)|$ for $l=0.5$ in the second
column of Fig.\ref{spin0510Fig}, finding that they look quite
similar to (and in fact, they are intermediate between) those in the
$l=0,1$ cases. Thus, at least for $l\leq 1$, our results suggest
that the wake picture described in \cite{FGM:2006bz} may also apply
to $R$-charged plasmas.

\begin{figure}[h]
\includegraphics[clip, scale=0.47, bb=47 198 525 788]{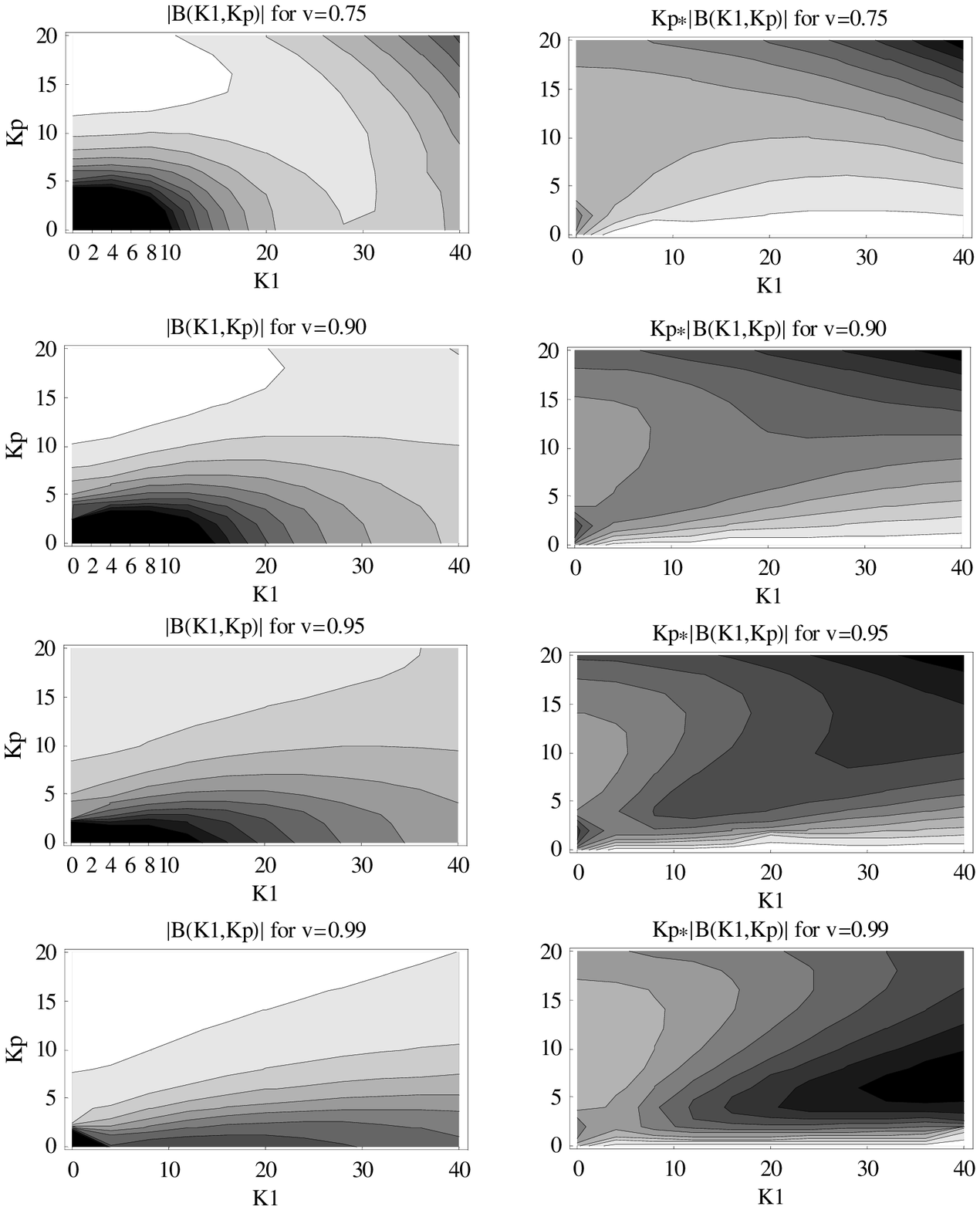}
\includegraphics[clip, scale=0.47, bb=47 198 525 788]{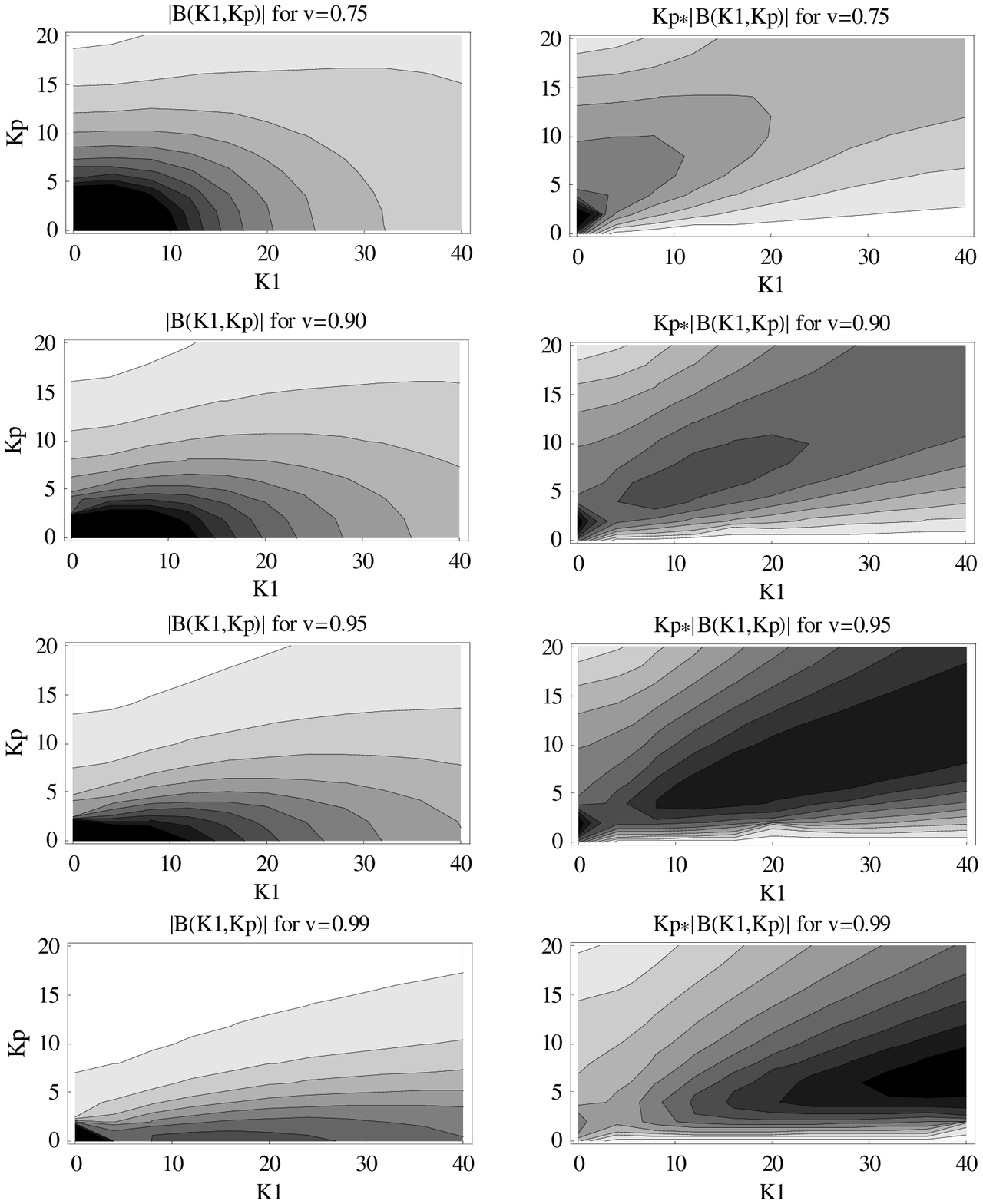}
\caption{Distributions of equi-value lines of $|B(K_1,K_\bot)|$
and $K_\bot|B(K_1,K_\bot)|$ in the plane $(K_{1},K_{\bot})$, for
$l=2$, at speeds $v=0.75,0.90,0.95,0.99$. The near field
contribution (\ref{nearFieldContribution}) has been subtracted
from $B(K_{1},K_{\bot})$. The left eight plots are computed using
the boundary conditions (\ref{bc}), while the right eight ones are
the corresponding plots with the new boundary conditions
$W_{NH}|_{r=1.001h_1}=0$, $W_{NB}|_{r=800h_1}=0$ imposed.}
\label{spin20Fig}
\end{figure}

What would happen if $l$ becomes larger? For a comparison, the plots of $|B(K_1,K_\bot)|$ and
$K_\bot|B(K_1,K_\bot)|$ for $l=2$ are displayed in the first and second columns of Fig.\ref{spin20Fig}. The pattern of such plots appears now to be somehow different from what we have seen in the case of $l\leq 1$. In particular, at relatively low speeds, the directionally peaked structure in $K_\bot|B(K_1,K_\bot)|$
seems no longer obvious. We should mention that in our numerical calculations,
plot patterns for $l$ being larger will depend more sensitively on how we
impose the boundary conditions. For instance, if we modify the second condition
in (\ref{bc}) by setting $W_{NB}=0$ at $r=800h_1$ instead of $700h_1$, then the plots for $l=2$ will have a new pattern, as shown in the third and forth columns of Fig.\ref{spin20Fig}, which is different from the previous one. Similar changes in plots are also observed at $l=0$, $0.5$ and $1$, but they are far less
remarkable, leaving the pattern qualitatively the same. Hence, our numerical results for $l=2$ are not quite reliable, which may
contain more errors than those in the $l\leq 1$ case.

Following \cite{FGM:2006bz}, we can determine the opening angle
$\theta$ between the velocity of the heavy quark and the directional peak
found in $K_{\bot}B(K_{1},K_{\bot})$. The results are
summarized in the following table: \beq{}
\begin{tabular}{l|cccc}
 $\theta$ & $v=0.75$ & $v=0.90$ & $v=0.95$ & $v=0.99$
 \cr\hline $l=0$& $0.59$ & $0.42$ & $0.28$ & $0.17$
 \cr\hline $l=0.5$ & $0.57$ &$0.39$ & $0.27$ & $0.16$
 \cr\hline $l=1$ & $0.55$ & $0.37$ & $0.26$& $0.15$
 \end{tabular}
\eeq We thus find a strong dependence of $\theta$ on $v$ in both the
neutral and charged cases. For fixed $v$, however, the opening angle
depends rather weakly on $l$ (at least in the region $l\leq 1$; as
we mentioned, our numerical results for $l=2$ are not quite
trustable), in a monotonically decreasing way.

In conclusions, we have studied the dissipative dynamics of a heavy quark passing through charged
${\cal N}=4$ SYM plasmas. Neutral plasmas were treated as a special case of the charged ones, where
we reproduced the main results of \cite{FGM:2006bz} using a different numerical method. Our results
for $l\leq h_{1}$ naively support the wake picture, but they are inconclusive for $l\sim 2h_{1}$ due
to numerical errors.

\end{document}